# Possible Nonlinear Anomalous Thermoelectric Effect in Organic Massive Dirac Fermion System

Toshihito Osada* and Andhika Kiswandhi

*Institute for Solid State Physics, University of Tokyo,*

*5-1-5 Kashiwanoha, Kashiwa, Chiba 277-8581, Japan.*

We propose a novel current-induced thermoelectric phenomenon, the nonlinear anomalous Ettingshausen effect (AEE), at zero magnetic field in inversion-asymmetric conductors. As an example, we discuss the weak charge ordering state in a layered organic conductor $\alpha$-(BEDT-TTF)$_2$I$_3$, which is a two-dimensional massive Dirac fermion system with a pair of tilted Dirac cones. The nonlinear AEE is a thermoelectric analogue of the nonlinear anomalous Hall effect, which is recently observed in $\alpha$-(BEDT-TTF)$_2$I$_3$, and these two effects generally appear simultaneously. The nonlinear AEE generates a transverse heat current, which exhibits rectifying characteristics, namely unidirectionality even under an AC current.

Recently, current-induced Berry curvature effects have been investigated in inversion-asymmetric conductors at zero magnetic field. In the equilibrium state with time reversal symmetry, these conductors do not exhibit any Berry curvature effect because the total Berry curvature summed up over occupied states is zero. However, in the current-carrying state, Berry curvature effects may be induced by a nonequilibrium distribution. The current-induced anomalous Hall effect (AHE), which is known as the nonlinear AHE, was originally proposed by Sodemann and Fu [1]. It was observed in thin films of Weyl semimetal $WTe_2$ [2, 3] and a bulk Dirac semimetal $Cd_3As_2$ [4], and also proposed in a chiral crystal of Te [5]. Current-induced orbital magnetization, which is a type of electromagnetic effect and known as the orbital Edelstein effect, was observed in strained monolayer $MoS_2$ [6, 7] and discussed in crystals with helical structures [8].

Furthermore, the nonlinear AHE was discussed and observed in an organic Dirac fermion (DF) system [9–11]. A layered organic conductor, $\alpha$-(BEDT-TTF)$_2$I$_3$, under high pressure is known as a two-dimensional (2D) massless DF system with a pair of tilted Dirac cones [12]. At ambient pressure, it undergoes a metal–insulator transition into a charge ordering (CO) insulating phase at $T_{CO}$ = 135 K, breaking the inversion symmetry. This CO phase is suppressed by applying a pressure and vanishes at approximately $P_c$ = 1.3 GPa. Above $P_c$, a metallic massless DF state is realized down to low temperatures. The DF state was originally demonstrated theoretically [13], and later confirmed via various experiments, such as those involving magnetotransport [14, 15], specific heat [16], and nuclear magnetic resonance [17]. In the massless DF state, the band dispersion around the Fermi level is regarded as a pair of tilted and anisotropic Dirac cones with Dirac points located at general points ($\mathbf{k}_0$ and $-\mathbf{k}_0$) in the $\mathbf{k}$ space. In this study, we focus on the "weak

CO" state immediately below $P_c$, where $T_{CO}$ is suppressed significantly. It has been suggested that the weak CO state is a massive DF state, where a small gap opens at the Dirac points [18-20]; in fact, this was experimentally confirmed recently [21]. The weak CO state provides an ideal platform for investigating the current-induced Berry curvature effects, although it generally comprises two types of CO domains [9]. In fact, nonlinear AHE has recently been observed in the weak CO state [11].

Herein, we propose a novel current-induced thermoelectric phenomenon, the nonlinear anomalous Ettingshausen effect (AEE), at zero magnetic field in inversion-asymmetric conductors. As a candidate system, we discuss the weak CO state of an organic multilayer DF system, $\alpha$-(BEDT-TTF)$_2$I$_3$.

As a simplified model of the weak CO state, we considered a 2D massive DF system comprising a pair of tilted Dirac cones ($\mathbf{k}_0$ and $-\mathbf{k}_0$ valleys) with a small gap. We employed the following effective Hamiltonian around the $s_v\mathbf{k}_0$ valley ($s_v = \pm 1$) [22, 23].

$$H^{(s_v)}(\mathbf{k}) = s_v \hbar v_F k_x \sigma_x + \hbar v_F k_y \sigma_y + s_v \hbar v_0 k_x \sigma_0 + \Delta \sigma_z . \tag{1}$$

Here, $\mathbf{k} = (k_x, k_y)$ is the 2D wave number measured from $s_v\mathbf{k}_0$. The matrices $\sigma_x$, $\sigma_y$, and $\sigma_z$ are Pauli matrices, and $\sigma_0$ is a $2 \times 2$ unit matrix. The $x$-axis is considered to be along the tilting direction of the Dirac cone, and $v_0$ indicates the magnitude of the tilting. In addition, we introduce a mass parameter $\Delta$, which breaks the inversion symmetry and opens the CO gap. Two types of CO domains have a mass parameter with opposite signs. The energy dispersion $E_\pm(\mathbf{k})$ and Berry curvature $\mathbf{\Omega}_\pm(\mathbf{k})$ of the conduction (+) and valence (−) bands around the $s_v\mathbf{k}_0$ valley can be obtained easily as follows:

$$E_\pm^{(s_v)}(\mathbf{k}) = s_v \hbar v_0 k_x \pm \sqrt{(\hbar v_F)^2 k_x^2 + (\hbar v_F)^2 k_y^2 + \Delta^2} , \tag{2}$$

$$\Omega_{\pm}^{(s_v)}(\mathbf{k}) = [\Omega_{\pm}^{(s_v)}(\mathbf{k})]_z \mathbf{n}_z = \frac{\mp s_v (\hbar v_F)^2 \Delta}{2\sqrt{(\hbar v_F)^2 k_x^2 + (\hbar v_F)^2 k_y^2 + \Delta^2}^3} \mathbf{n}_z. \qquad (3)$$

Here, $\mathbf{n}_z$ is a unit vector along the stacking ($z$) axis. It is noteworthy that the Berry curvature does not depend on the tilting. We assume a small imbalance between the electron and hole densities ($n - p > 0$). $E_{\pm}(\mathbf{k})$ and $[\Omega_+(\mathbf{k})]_z$ are schematically illustrated in Fig. 1.

In the equilibrium state, the total Berry curvature summed up over occupied states is vanished because the sign of $[\Omega_{\pm}(\mathbf{k})]_z$ is opposite in two valleys, resulting in no Berry curvature effects. However, Berry curvature effects may appear in the current-carrying state with a nonequilibrium distribution, wherein the in-plane electric field $\mathbf{E}$ is applied and a stationary electric current is flowing [1].

First, we briefly review the nonlinear AHE in the present model [9, 10]. The current-induced nonequilibrium distribution breaks the balance of the anomalous velocity $(e/\hbar)\mathbf{E} \times \Omega(\mathbf{k})$ between two valleys [24], causing a nonlinear AHE. The anomalous Hall current in the current-carrying state is represented as follows:

$$\mathbf{j}^{(2)} = \frac{e^3 \tau}{\hbar^2} \{(\Lambda_+^z + \Lambda_-^z) \cdot \mathbf{E}\}(\mathbf{E} \times \mathbf{n}_z), \qquad (4)$$

$$\Lambda_{\pm}^z \equiv \frac{1}{(2\pi)^3} \iiint \frac{\partial [\Omega_{\pm}(\mathbf{k})]_z}{\partial \mathbf{k}} f_{\pm}^0(\mathbf{k}) d\mathbf{k}^3. \qquad (5)$$

Here, $\Lambda_{\pm}^z$ is the Berry curvature dipole (BCD). The $\alpha$-component of $\mathbf{j}^{(2)}$ is written as $j_\alpha^{(2)} = \chi_{\alpha\beta\gamma} E_\beta E_\gamma$ using Einstein's notation. The nonlinear Hall conductivity $\chi_{\alpha\beta\gamma}$ is expressed by $\chi_{\alpha\beta\gamma} = (e^3 \tau / \hbar^2)\varepsilon_{\alpha\gamma z}[\Lambda_+^z + \Lambda_-^z]_\beta$ using the Levi–Civita symbol $\varepsilon_{\alpha\gamma z}$. In this model, the non-zero elements of $\chi_{\alpha\beta\gamma}$ are only $\chi_{yxx}$ and $\chi_{xxy} = -\chi_{yxx}$; therefore, the nonlinear AHE is represented as $\mathbf{j}^{(2)} = \left(\chi_{xxy} E_x E_y\right)\mathbf{n}_x + \left(\chi_{yxx} E_x^2\right)\mathbf{n}_y = \chi_{yxx} E_x \left(\mathbf{n}_z \times \mathbf{E}\right)$, where $\mathbf{n}_x$ and $\mathbf{n}_y$ are unit vectors in the *x*- and *y*-directions, respectively.

Next, we discuss a new current-induced thermoelectric effect, the nonlinear AEE, in the similar way as the nonlinear AHE, in the present model. Generally, the Berry curvature correction of the heat current under an electric field is written as follows [25]:

$$\mathbf{j}_Q = \mathbf{E} \times \frac{1}{(2\pi)^3} \iiint \frac{e}{\hbar} \mathbf{\Omega}(\mathbf{k}) \left[ \{E(\mathbf{k}) - \mu\} f^0(\mathbf{k}) + k_B T \log\left\{1 + e^{-(E(\mathbf{k}) - \mu)/k_B T}\right\} \right] d\mathbf{k}^3. \quad (6)$$

Here, $f^0(\mathbf{k}) = 1/[\exp\{(E(\mathbf{k}) - \mu)/k_B T\} + 1]$ is the equilibrium distribution with a chemical potential $\mu$. The first term of the integral reflects the contribution from the anomalous velocity of wave packets, whereas the second term corresponds to that from the itinerant bulk current such as the edge current. Because $\mathbf{j}_Q$ is perpendicular to $\mathbf{E}$, a finite $\mathbf{j}_Q$ corresponds to the AEE at zero magnetic field, which is a thermoelectric analogue of the AHE. The AEE is associated with the anomalous Nernst effect (ANE) by the Onsager relation [25]. Under time reversal symmetry, $\mathbf{j}_Q$ vanishes because $\mathbf{\Omega}(-\mathbf{k}) = -\mathbf{\Omega}(\mathbf{k})$. In fact, in the equilibrium state of the present model with time reversal symmetry, the AHE and AEE cannot be expected, resulting from the cancellation between the two valleys.

However, in the current-carrying state, the AHE and AEE are revived as the nonlinear AHE and nonlinear AEE, respectively. According to Boltzmann theory, in the current-carrying state, the distribution function is shifted by $\Delta f(\mathbf{k}) = \{-\partial f^0(\mathbf{k})/\partial E\} \tau \mathbf{v}(\mathbf{k}) \cdot (-e)\mathbf{E}$ under uniform temperature and chemical potential. Here, $\tau$ is a constant relaxation time and $\mathbf{v}(\mathbf{k}) = (1/\hbar)\partial E(\mathbf{k})/\partial \mathbf{k}$ is the group velocity of wave packets. This distribution change causes a nontrivial heat current as follows:

$$\mathbf{j}_Q^{(2)} = \mathbf{E} \times \frac{1}{(2\pi)^3} \iiint \frac{e}{\hbar} \mathbf{\Omega}(\mathbf{k})\{E(\mathbf{k}) - \mu\}\Delta f(\mathbf{k})d\mathbf{k}^3 . \tag{7}$$

Suppose that the electric field $\mathbf{E}$ is applied along the $x$-axis (tilting axis of the Dirac cone). As illustrated in Fig. 1, the distribution is modified by $\mathbf{E}$, and the occupied region is shifted from the Fermi surface to the $-x$ direction in both valleys. If the Dirac cones are tilted with a finite $v_0$, then the shift in the occupied states breaks the cancellation between the sums of $[\mathbf{\Omega}_+^{(+)}(\mathbf{k})]_z$ and $[\mathbf{\Omega}_+^{(-)}(\mathbf{k})]_z$ over the occupied states. This causes a nonlinear AEE. The anomalous heat current in the current-carrying state is represented as follows:

$$\mathbf{j}_Q^{(2)} = -k_B T \frac{e^3 \tau}{\hbar^2} \left\{ \left(\mathbf{\Theta}_+^z + \mathbf{\Theta}_-^z\right) \cdot \mathbf{E} \right\}(\mathbf{E} \times \mathbf{n}_z), \tag{8}$$

$$\mathbf{\Theta}_\pm^z \equiv \frac{1}{(2\pi)^3 k_B T} \iiint \frac{\partial \{E_\pm(\mathbf{k}) - \mu\}[\mathbf{\Omega}_\pm(\mathbf{k})]_z}{\partial \mathbf{k}} f^0(\mathbf{k})d\mathbf{k}^3 . \tag{9}$$

Here, $\mathbf{\Theta}_\pm^z$ is a thermoelectric analogue of the BCD and is referred to as the thermoelectric BCD herein. The $\alpha$-component of $\mathbf{j}_Q^{(2)}$ is written as $[\mathbf{j}_Q^{(2)}]_\alpha = \chi_{\alpha\beta\gamma}^Q E_\beta E_\gamma$. The nonlinear Ettingshausen coefficient $\chi_{\alpha\beta\gamma}^Q$ is expressed as $\chi_{\alpha\beta\gamma}^Q = -k_B T(e^2\tau/\hbar^2)\varepsilon_{\alpha\gamma z}[\mathbf{\Theta}_+^z + \mathbf{\Theta}_-^z]_\beta$. The non-zero elements of this coefficient are only $\chi_{yxx}^Q$ and $\chi_{xxy}^Q = -\chi_{yxx}^Q$ because $[\mathbf{\Theta}_\pm^z]_y = 0$ in the present model. The nonlinear AEE originates from the Berry curvature of the system, but not from the group velocity; therefore, it is not directly related to the normal linear thermoelectric nature of the system [26].

Here, we should note that the nonlinear AEE was derived from the linear Boltzmann transport theory, in which the distribution shift $\Delta f(\mathbf{k})$ is given by the first order of $\mathbf{E}$ and $\tau$. The second order solution of Boltzmann equation gives a trivial nonlinear (second order) heat current parallel to $\mathbf{E}$ or the third order Ettingshausen heat current perpendicular to **E**. Therefore, the nonlinear AEE derived here gives the lowest order contribution to the perpendicular component of **j**$_Q$. This is the same approximation as the original derivation of the nonlinear AHE [1].

We evaluated the thermoelectric BCD for a 2D tilted massive Dirac fermion system. The *y*-component of the BCD $[\Theta^z_\pm]_y$ is always zero because the integrand of (4) is an odd function of $k_y$. As for the *x*-component $[\Theta^z_\pm]_x$, the two valleys provide equal contributions. Figure 2(a) shows the chemical potential dependence of $[\Theta^z_+ + \Theta^z_-]_x$, which is the sum of the thermoelectric BCDs of the conduction and valence bands at a fixed tilting, $v_0/v_F = 0.8$. Here, $[\Theta^z_+ + \Theta^z_-]_x$ is normalized by $4\hbar v_F/(2\pi)^2 c\Delta$ including spin degeneracy, where $c$ is the interlayer distance. The thermoelectric BCD $[\Theta^z_\pm]_x$ vanishes when the Dirac cones have no tilting ($v_0/v_F = 0$). The shaded region around $\mu/|\Delta| = 0$ corresponds to the energy gap between the conduction and valence bands. At finite temperatures, $[\Theta^z_+ + \Theta^z_-]_x$ has a finite value even when $\mu$ is located in the energy gap, because thermally excited carriers contribute to the thermoelectric BCD. This implies that the thermoelectric BCD can appear even in the insulating state.

In $\alpha$-(BEDT-TTF)$_2$I$_3$, the band parameters were estimated as $v_F = 1.0 \times 10^5$ m/s and $v_0 = 0.8 \times 10^5$ m/s from a comparison with first principles calculation [23]. The interlayer distance was reported as $c = 1.75$ nm [27]. Using these values, we can quantitatively estimate the nonlinear AEE coefficient $\chi^Q_{yxx} = [\mathbf{j}^{(2)}_Q]_y/E_x^2$ in $\alpha$-(BEDT-TTF)$_2$I$_3$. Figure 2(b) shows the carrier density dependence of $\chi^Q_{yxx}$, which was normalized by the mass parameter $\Delta$ and the scattering relaxation time $\tau$. The carrier density imbalance, $n - p$, is the difference in the densities between thermally excited electrons and holes. Although the nonlinear AHE coefficient $\chi_{yxx}$ is an odd function of $n - p$ [9], $\chi_{yxx}^Q$ is an even function of $n - p$ because the heat current of the electrons and holes flows in the same direction.

Actual $\alpha$-(BEDT-TTF)$_2$I$_3$ crystals are slightly electron-doped, likely due to the partial lack of I$_3^-$ ions. They demonstrate a finite Hall effect, suggesting an uncompensated carrier density $n - p$ on the order of $10^{15} - 10^{16}$ cm$^{-3}$ at low temperatures [12]. The mass parameter was estimated from $|\Delta| \sim k_B T_{CO}$. In a previous study, the scattering broadening of the $n = 0$ Landau level was experimentally estimated to be 3 K [28], which corresponds to $\tau \sim 2.5$ ps if the scattering time does not change under the magnetic field. Assuming realistic parameters, i.e., $n - p = 10^{16}$ cm$^{-3}$, $\tau = 1$ ps, and $k_B T / |\Delta| = 0.5$, we can obtain $\chi_{yxx}^Q = [\mathbf{j}_Q^{(2)}]_y / E_x^2 = -0.45$ mW/V$^2$ from Fig. 2(b).

The nonlinear AEE in the present system with Dirac cones tilting in the $x$-direction is represented as $\mathbf{j}_Q^{(2)} = \left(\chi_{xxy}^Q E_x E_y\right)\mathbf{n}_x + \left(\chi_{yxx}^Q E_x^2\right)\mathbf{n}_y = \chi_{yxx}^Q E_x \left(\mathbf{n}_z \times \mathbf{E}\right)$. Figure 3(a) illustrates the schematic configuration, and Fig. 3(b) shows the dependence of the $\mathbf{j}_Q^{(2)}$ vector on the electric field direction measured by the azimuthal angle $\theta$ in the case involving $\mu > 0$, $v_0 > 0$, and $\Delta > 0$. The direction of $\mathbf{j}_Q^{(2)}$ is always perpendicular to the electric field, and the trajectory of the end point of $\mathbf{j}_Q^{(2)}$ is a circle passing the origin. It is noteworthy that the $y$-component $[\mathbf{j}_Q^{(2)}]_y = \chi_{yxx}^Q E_x^2$ is always negative ($\chi_{yxx}^Q < 0$); in other words, the heat current always flows in the $-y$-direction under electric fields of any direction. $\mathbf{j}_Q^{(2)}$ never changes when $\mathbf{E}$ is reversed. This rectifying characteristic is the most remarkable feature of the nonlinear AEE which is the second-order response to the electric field.

In the current-carrying state, conventional linear thermoelectric effects, i.e., the Seebeck and Peltier effects, also occur at zero magnetic field. Moreover, the system must exhibit the nonlinear AHE and nonlinear ANE, which is a counterpart of nonlinear AEE. To consider the voltage and temperature response to the charge current, we must simultaneously consider the charge and heat current at zero magnetic field.

$$\begin{cases} \mathbf{j} = \sigma \mathbf{E} + \chi E_x(\mathbf{n}_z \times \mathbf{E}) + (\sigma S)(-\nabla T) + \frac{\chi^Q}{T}(-\nabla_x T)(\mathbf{n}_z \times \mathbf{E}) \\ \mathbf{j}_Q = T(\sigma S)\mathbf{E} + \chi^Q E_x(\mathbf{n}_z \times \mathbf{E}) + \kappa(-\nabla T) \end{cases}. \tag{10}$$

Here, $\sigma$, $\kappa$, and $S$ denote the electric conductivity, electron thermal conductivity, and Seebeck coefficient, respectively. $S$ is negative for electrons ($\mu > 0$). For simplicity, we assumed that these parameters are scalar quantities. The first and third terms of $\mathbf{j}$ correspond to contributions from Ohm's law and the Seebeck effect, respectively, whereas the first and third terms of $\mathbf{j}_Q$ correspond to contributions from the Peltier effect and Fourier's law, respectively. The second term of $\mathbf{j}$ represents the nonlinear AHE with $\chi \equiv \chi_{yxx} = -\chi_{xxy}$, whereas the second term of $\mathbf{j}_Q$ represents the nonlinear AEE with $\chi^Q \equiv \chi^Q_{yxx} = -\chi^Q_{xxy}$. It is noteworthy that $\chi^Q$ is almost negative but $\chi$ is positive in the case involving $\mu > 0$, $v_0 > 0$, and $\Delta > 0$. The fourth term of $\mathbf{j}$ represents the nonlinear ANE, which originates from the breaking of Berry curvature balance between two valleys due to the distribution change under the temperature gradient.

We consider a thermally isolated system in which the charge current $\mathbf{j}$ flows along the direction tilted by the angle $\alpha$ from the $x$-axis (tilting axis of Dirac cones) as shown in Fig. 3(c), in which the components of **j** and **j**$_Q$ in Eq. (10) are schematically indicated by dashed and solid arrows, respectively. The total heat current is zero ($\mathbf{j}_Q = 0$) because of thermal isolation. We define the "$\parallel$"- and "$\perp$"-axes which are rotated by $\alpha$ from the *x*- and *y*-axes, respectively. Hence, **j** is parallel to the $\parallel$-axis with $j_\parallel = \mathbf{j} \cdot \mathbf{n}_\parallel$ ($\mathbf{n}_\parallel$ is a unit vector along the $\parallel$-axes). These conditions correspond to the measurements in the shaded rectangular sample in Fig. 3(c). By solving (9) under the condition $\mathbf{j}_Q = 0$ and $(\chi / \sigma^2) j_\parallel \ll 1$, we can obtain the following formulae for the lowest order contribution of $j_\parallel$ to the voltage and temperature gradient.

$$\begin{aligned}
E_\parallel &\simeq \frac{1}{\sigma(1-ZT)} j_\parallel, \\
E_\perp &\simeq -\frac{(\chi - \sigma S \chi^Q / \kappa) \cos\alpha}{\sigma^3 (1-ZT)^3} j_\parallel^2, \\
-\nabla_\parallel T &\simeq -\frac{TS}{\kappa(1-ZT)} j_\parallel, \\
-\nabla_\perp T &\simeq -\frac{(\chi^Q - TS\chi) \cos\alpha}{\kappa \sigma^2 (1-ZT)^3} j_\parallel^2.
\end{aligned} \quad (11)$$

Here, $ZT \equiv (S^2 \sigma / \kappa) T$ is the thermoelectric figure of merit. It is noteworthy that the fourth term of **j** (the nonlinear ANE) does not contribute to (11) because it gives higher order contributions of $j_\parallel$ under the boundary condition $\mathbf{j}_Q = 0$. Furthermore, in the case of $|S| \ll k_B / e$ (~ 86.2 $\mu$V/K), which is satisfied in $\alpha$–(BEDT-TTF)$_2$I$_3$ with $|S|$ ~ 6 $\mu$V/K at low temperatures [26], (11) can be simplified to $E_\parallel \sim (1/\sigma) j_\parallel$, $E_\perp \sim -(\chi \cos\alpha / \sigma^3) j_\parallel^2$, $-\nabla_\parallel T \sim -(TS/\kappa) j_\parallel$, and $-\nabla_\perp T \sim -(\chi^Q \cos\alpha / \kappa \sigma^2) j_\parallel^2$. Between the side edges separated by the distance of $L_\perp$ in Fig. 3(c), the anomalous Hall voltage $E_\perp L_\perp$ and anomalous temperature difference $(-\nabla_\perp T)L_\perp$ are induced by the current, owing to the nonlinear AHE and the nonlinear AEE, respectively. They are characterized by the square dependence on the electric current, and their amplitude depends on the angle $\alpha$. These features are depicted in Fig. 3(d). The heat current always flows from the left side to the right side of the Dirac cone tilting axis regardless of the current direction, resulting in the cooling (heating) of the left (right) surface. This rectifying characteristic might enable thermoelectric cooling using AC current or illumination of electromagnetic waves.

The voltage and temperature differences satisfy the following relation.

$$\frac{-\nabla_\perp T}{E_\perp} \simeq \frac{\sigma}{\kappa}\frac{\chi^Q}{\chi} = -\frac{3e}{\pi^2 k_B}\frac{[\Theta_+^z + \Theta_-^z]_x}{[\Lambda_+^z + \Lambda_-^z]_x} = -3.53\ [\text{K/mV}]\ \frac{[\Theta_+^z + \Theta_-^z]_x}{[\Lambda_+^z + \Lambda_-^z]_x}. \tag{12}$$

Here, we used the Wiedemann–Franz law $\kappa/\sigma T = (\pi^2/3)(k_B/e)^2$. $[\Theta_+^z + \Theta_-^z]_x$ and $[\Lambda_+^z + \Lambda_-^z]_x$ are shown in Fig. 2(a) herein and Fig. 2(b) of Ref. 9, respectively. The ratio $[\Theta_+^z + \Theta_-^z]_x / [\Lambda_+^z + \Lambda_-^z]_x$ is generally of the order of 1, but it can exceed 10 in the gap ($|\mu| < |\Delta|$) at finite temperatures because $[\Theta_+^z + \Theta_-^z]_x$ remains finite but $[\Lambda_+^z + \Lambda_-^z]_x$ approaches zero. Experimentally, the anomalous Hall voltage of $E_\perp L_\perp \simeq 40\ \mu\text{V}$ was observed in the weak CO state in $\alpha$-(BEDT-TTF)$_2$I$_3$ [11]; therefore, the temperature difference $(-\nabla_\perp T)L_\perp$ of the order of 0.1 K is expected to be induced simultaneously between voltage contacts. This value is small but may be within the observable range.

Finally, we comment on the effect of multiple domains in the crystal. In the actual $\alpha$-(BEDT-TTF)$_2$I$_3$ crystal, two types of CO domains with opposite signs of $\Delta$ appear after the CO transition. In a system with multiple domains, the nonlinear AHE and nonlinear AEE tend to be cancelled between the two types of domains. To reduce this cancellation, we proposed the current-field-cooling method, which enhances the selective formation of a single type of domain using the orbital Edelstein effect [9].

In conclusion, we discussed a new current-induced thermoelectric phenomenon at zero magnetic field, i.e., the nonlinear AEE, considering the weak CO state of $\alpha$-(BEDT-TTF)$_2$I$_3$ as an example. The nonlinear AEE, which is a thermoelectric analogue of the nonlinear AHE, is caused by the finite thermoelectric BCD of the system. The nonlinear AEE generates a transverse heat current, which exhibits rectifying characteristics, namely unidirectional flow even under an AC current. The nonlinear AEE appears simultaneously with the nonlinear AHE, and it is in the observable range in $\alpha$-(BEDT-TTF)$_2$I$_3$. The nonlinear AEE is expected also in other 2D systems showing the nonlinear AHE, such as few layer WTe$_2$ [2, 3].

**Acknowledgements**

The authors thank Dr. T. Taen, Dr. K. Uchida, and Dr. M. Sato for their valuable input. This work was supported by JSPS KAKENHI Grant No. JP20H01860.

# References

*corresponding author, osada@issp.u-tokyo.ac.jp

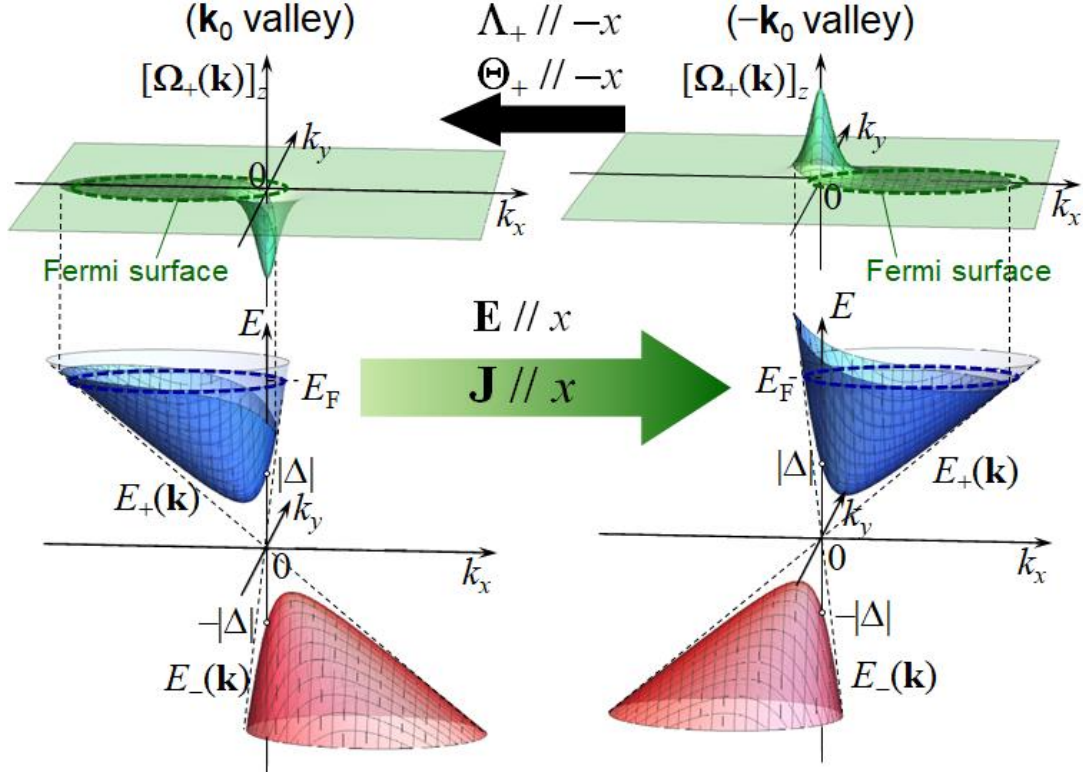

**Figure 1** (color online)

Electron occupation around $\mathbf{k}_0$ and $-\mathbf{k}_0$ valleys in current-carrying state of 2D massive DF system with tilted Dirac cones. Upper and lower panels illustrate Berry curvature of conduction band and band dispersion with a small gap, respectively. Nonequilibrium distribution under electric field $\mathbf{E} \parallel x$ causes a shift in the occupied state in $\mathbf{k}$-space, breaking the balance of the Berry curvature of the occupied states between the two valleys.

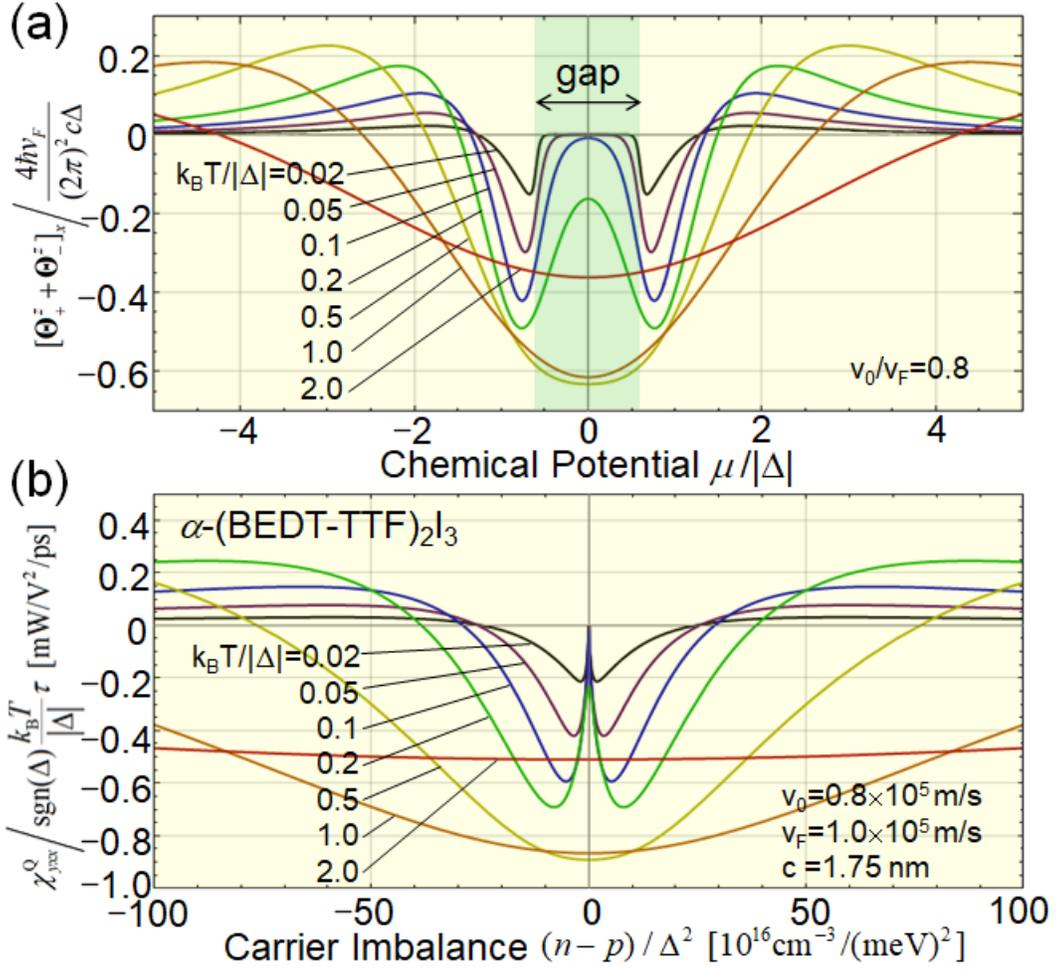

**Figure 2** (color online)

(a) Sum of thermoelectric BCD of conduction and valence bands as a function of chemical potential at several temperatures for Dirac cone tilting $v_0/v_F = 0.8$. (b) Response coefficient $\chi^Q_{yxx} = [\mathbf{j}^{(2)}_Q]_y / E_x^2$ as a function of difference between electron and hole densities $n-p$, estimated using parameters of $\alpha$-(BEDT-TTF)$_2$I$_3$ ( $c = 1.75$ nm, $v_F = 1.0 \times 10^5$ m/s, and $v_0 = 0.8 \times 10^5$ m/s ).

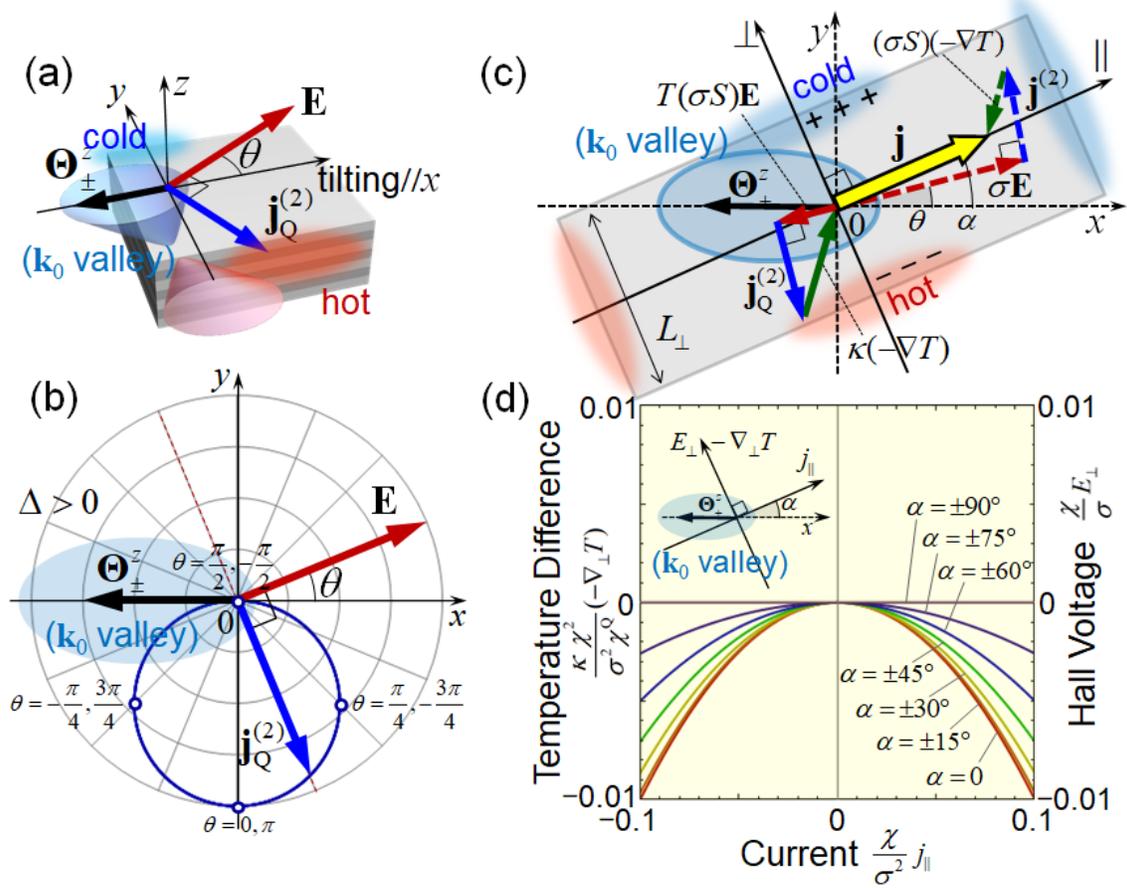

**Figure 3** (color online)

(a) Configuration of nonlinear anomalous Ettingshausen effect. (b) Trajectory of nonlinear heat current vector $\mathbf{j}_Q^{(2)}$ when direction $\theta$ of in-plane electric field $\mathbf{E}$ is changed. (c) Configuration of elements of electric current $\mathbf{j}$ (dashed arrows) and heat current $\mathbf{j}_Q$ (=0) (solid arrows) when linear and nonlinear components of electric and thermal transport coexist. (d) Current dependence of temperature difference due to nonlinear anomalous Ettingshausen effect and voltage drop due to nonlinear anomalous Hall effect in the case of $|S| \ll k_B/e$. Both show the rectifying characteristics.

17